# What makes you popular: beauty, personality or intelligence?

Fronzetti Colladon, A., Grippa, F., Battistoni, E., Gloor, P. A., & La Bella, A.

# What Makes You Popular: Beauty, Personality or Intelligence?

Fronzetti Colladon, A., Grippa, F., Battistoni, E., Gloor, P. A., & La Bella, A

**Abstract**

This study explores the determinants of popularity within friendship and advice networks. We involved almost two hundreds college students in an experiment to predict how personality traits, self-monitoring, creativity, intelligence, energy, and beauty influence the development of friendship and advice networks. Our results indicate that physical attractiveness is key to develop both friendship and task-related interactions, whereas perceived intelligence and creativity play an important role in the advice network. Our findings seem to support the idea that there might be a kernel of truth in the stereotype that attractiveness correlates with positive social traits and successful outcomes.

# 1. Introduction

In this study we are interested in finding the main determinants of popularity in social contexts where individuals rely on each other to increase their knowledge and develop meaningful relationships. We define popularity as the ability to attract peers with the purpose to establish friendship ties and create a network of people to rely on for advice. Network researchers have been mainly focused on studying the consequences of network centrality rather than the antecedents (Allen et al., 2016; Gloor et al., 2016; Gloor et al., 2017; Klein et al., 2004; Mehra et al., 2001). Furthermore, prior research focused on studying the possible determinants of centrality using a singular approach. On the contrary, here we combine and extend the contribution of previous work in the field of personality traits (Fang et al., 2015; Klein et al., 2004; Mehra et al., 2001; Pollett et al., 2011); creativity (Cattani and Ferriani, 2008; Fleming and Marx, 2006), intelligence (Jeung, 2013) and beauty (Eagly et al., 1991).

In this study, we focus on two types of informal networks: advice-seeking networks and friendship networks. Studying the properties of the advice networks is key to understand how information and knowledge is exchanged; at the same time, we concentrate on the informal network of friendship since factors such as trust, familiarity and confidence and can impact the likelihood to ask for support and advice. A higher popularity in the friendship and advice network could be beneficial for the entrepreneurs' efforts to establish new contacts and increase business opportunities. In particular, entrepreneurs who are central in the advice network have access to information coming from multiple sources, which helps them leverage weak ties and unleash the creative process that is conducive to innovation. Researchers have long been interested in the relation between entrepreneurship and creativity (Whiting, 1988). Recent studies have emphasized the benefits of bringing many areas of specialized knowledge together to promote creativity (e.g., Battistoni and Fronzetti Colladon, 2014; Demsetz, 1988). Allen et al. (2016) demonstrated that a more dynamic communication style and more diverse social ties are beneficial to innovation. Another recent study found that being a very popular entrepreneur can help make their product strongly desired by customers, due to an increased presence in both traditional and digital media outlets and an increased brand awareness (Battistoni et al. 2013). Obviously, the positive correlation between brand value and popularity is not confirmed when the entrepreneur's popularity is based on negative events.

In order to understand the factors determining individual popularity we use cognitive, psychological as well as sociological theories to explore what causes individuals' interaction choice. Fewer studies have been investigating how individuals acquire central positions within their social networks. In this paper we investigate which combination of factors, from personality traits – such as extraversion or self-monitoring – to cognitive processes – such as creativity or intelligence – has a higher impact on social network positions.

We focus on the informal relationships between individuals and how they are shaped by personality traits, cognitive processes and other individual attributes. Through these informal connections, individuals build relationships that strengthen their ability to be successful and reach their goals. Benefits of key social positions in advice networks were extensively investigated in past research (e.g. Battistoni and Fronzetti Colladon, 2014). Cattani and Ferriani (2008) provided evidence of the advantages coming from the ability to be in touch with the network core without losing touch with the periphery: by holding a central position in their informal social networks, individuals are more likely "*to acquire knowledge without acquiring the ties that typically bind such knowledge to particular worlds*" (Hargadon, 2005, p. 17).

The paper is organized as follows. In the next section, we review the literature on the main determinants of popularity and then we present the hypotheses (§2). After describing how data was collected and variables measured (§3), we present the results of the analyses (§4). We conclude by discussing the main implications of the findings, the limitations of our study, and important topics for future research (§5 and §6).

## 2. Homophily and Determinants of Popularity

Psychological studies in the area of social cognition (Fiske et al., 2006) suggest that human social perception is influenced by two main factors: liking, which captures traits like friendliness and sociability; and respecting, which captures factors like intelligence, competence, and efficiency. More recent research gave evidence to the possible interdependence of friendship and advice-related interactions between individuals (Kilduff, 1990; Snijders et al., 2013). For example, Snijders et al. (2013) suggested that in the advice network homophily, defined as the individuals' tendency to choose to connect with others exhibiting similar attributes, is mediated by friendship relations which are influenced by mutual attraction. Similarly, in our study we focus on the co-evolution of friendship and advice relationships, taking into account the way they influence one another.

We define the advice network as a set of "*relations through which individuals share resources such as information, assistance, and guidance*" (Sparrowe et al., 2001, p.317). The friendship network is defined by the ties of affection and companionship that connect individuals (Baldwin et al., 1997). We rely on the methodologies and techniques of social network analysis to operationalize individual's centrality within a network (Wasserman and Faust, 1994). There are, of course, different dynamics at play when it comes to attract friends or advisees. Actors with a high in-degree centrality in the advice network are usually preferred for their input on a specific content/competence area, while actors who have a high in-degree centrality in the friendship network are chosen for their companionship (Klein et al., 2004).

Recent studies on networks and social contagion (Christakis and Fowler, 2013; Aral et al., 2013) demonstrated the powerful effect of homophily on several individual characteristics such as age and gender. It has been widely demonstrated that homophily can help identify the emergence of friendship or advice networks based on some traits similarity. Homophily is one of the elements that can significantly affect popularity, since being similar and part of a more cohesive social cluster could lead to be more popular, in the sense of collecting more incoming social ties. For this reason we control for homophily effects while studying popularity. In this paper, we aim to understand the influence of less explored traits influencing people's attitude to "flock together", such as beauty, creativity, and intelligence (McPherson et al., 2001), as well as to explore the direct influence of these traits on popularity, both in terms of friendship and advice-seeking. The following paragraphs discuss in detail the main determinants of popularity that we consider important to explain a central position in both the advice and friendship networks, as well as the connected homophily effects: beauty, personality traits, performance, energizing ability, intelligence and creativity.

### 2.1. The Attractiveness Factor

The definition of physical attractiveness is still controversial and researchers have often defined people as being attractive based on raters' agreement on their attractiveness.

Attractiveness is a quality of individuals that is often quantified in terms of the consensus among opinions of others. Langlois et al. (2000) conducted a meta-analysis that showed how attractiveness is a plus in a variety of important real-life situations, both within and across cultures. Attractiveness seems to be strongly related to popularity, and moderately related to both intelligence and performance. People seem to agree about who is attractive and who is not, independently from gender and age differences (Langlois et al., 2000; Eagly et al., 1991; Krantz, 1987). Social psychologists have extensively demonstrated that individuals are attracted by other people who are similar to them, or by people who are naturally attractive, either in their appearance or personality (Byrne et al., 1968; Peskin and Newell, 2004; Verhulst et al., 2010). Verhulst et al. (2010, p. 116) noted how "*preconscious familiarity likely activates positive feelings that are used to construct subsequent evaluative judgments, including those related to competence*". Sources of personal likability such as face similarity and familiarity can contribute to the formation of an informal network and are a powerful way to generate positive feelings (Casciaro and Lobo, 2008). It is well established in literature that competence, affect-based trust and social interaction impact task-related interactions (Hinds et al., 2000). What seems to be less investigated is the influence of perceived beauty on the ability of people to develop and maintain central positions in their social networks.

There are several empirical studies suggesting that people expect physically attractive others to be more intelligent than less attractive others (Jackson et al., 1995; Zebrowitz et al., 2002). Other studies reinforce this physical attractiveness stereotype demonstrating that people who are considered by others as physically attractive are also perceived as being more popular and social competent (Eagly et al., 1991). Building on Berger's (1972) work, Webster and Driskell (1983) demonstrate that beauty produces ideas of task competencies: people who possess the high state of attractiveness are also expected to possess the high state of other specific characteristics (e.g. piloting a plane), and general, unlimited characteristics.

There has been an interesting exchange of ideas and empirical evidences amongst scholars over the question whether beautiful people are also more intelligent and perform better than others. In response to the study by Kanazawa and Kovar (2004) - stating that there is a positive correlation between beauty and intelligence - other scholars followed with empirical research and counter-arguments to reject those results (Denny, 2008). In this paper, we are not interested in studying the link between physical attractiveness and intelligence or performance. Our goal is to understand whether being perceived as more attractive, or beautiful, has an impact on the ability to develop stronger friendship and advice networks.

**2.2. The Impact of Personality**

Based on previous research following the similarity/attraction theory (Byrne, 1971; Krebs and Adinolfi, 1975), people tend to be more attracted to others who share their attitudes and personality traits. In their study, Casciaro and Lobo (2008) found that individuals "*consistently showed a preference for people they liked but considered mediocre at the task over competent but unpleasant people*" (p. 675). Accordingly, we expect a significant homophily effect coming from similar personality traits.

Both the five personality traits and self-monitoring have attracted the attention of several network researchers from a theoretical and empirical perspective (Fang et al., 2015; Klein et al., 2004; Mehra et al., 2001; Pollett et al., 2011). For example, Fang et al. (2015) recently found that individuals' personality traits and their social network positions both matter for job performance and career success.

In this paper we extend these recent empirical efforts by including other determinants such as creativity, intelligence, energy and beauty. Recently, Battistoni and Fronzetti Colladon (2014) focused on the relation between personality traits and social network positions of college students with regard to advice networks. Their study provided empirical evidence of significant associations between key network positions and traits such as conscientiousness, neuroticism and agreeableness, but found no evidence for a relationship with extraversion or openness to experience.

Recent studies suggest that people perform better when they demonstrate personality traits such as self-monitoring or conscientiousness, which are highly respected in the work environment (Fang et al., 2015; Barrick et al., 2001). For example, Furnham and Zhang (2005) found that factors - such as fluid intelligence - and personality traits - such as conscientiousness - could be predictors of academic performance.

Kalish and Robins (2006) suggest that psychological predispositions are also important factors in the formation of ego-networks. For example, Hallinan and Kubitschek (1988) analyze students' friendship networks and explain how those with a high in-degree centrality are more intolerant to intransitive triads and tend to remove intransitivity over time.

Individuals who exhibit extraversion tend to be sociable, assertive, positive and active, while neuroticism has been empirically associated to a lower level of sociability and to experiencing anxiety and a sense of vulnerability (Costa and McCrae, 1992; Russell et al., 1997). Agreeableness is a trait that is often associated to trust in other individuals, honest communication, altruistic and cooperative behavior, as well as sympathetic attitudes. Conscientiousness relates to the degree to which individuals are methodical, disciplined, prefer order and structure, while openness to experience refers to the extent to which individuals are open to novel actions, ideas, and values (Costa and McCrae, 1992). Kalish and Robins (2006) showed that individuals characterized by a high level of extraversion are likely to stay close to their social partners and connect them with other people. On the other hand, they found that neurotic individuals do not maintain close relationships with their social partners. These and similar studies are mainly focused on the impact of personality on the quality and structure of an individual's social network; they do not tell us, though, which traits are the key determinants of centrality in informal social networks.

Eisenkraft and Eifenbein (2010) investigated the impact of these personality types on social relationship and found that extraverted individuals might induce negative emotions in the people with whom they interact, probably because of the extraverted individuals' tendency to seem dominant which could undermine the sense of power in other people. They also found that individuals who were more central in their social networks were often inspiring positive emotions in their colleagues. Klein et al. (2004) found that individuals who are low in neuroticism tend to have high degree centrality scores in advice and friendship networks. A recent study on how friendship networks develop in late adolescence suggested that individuals high on agreeableness tended to be selected more as friends, and those high on extraversion tended to select more friends than those low on this trait. The study also found that individuals tended to select friends with similar levels of agreeableness, extraversion, and openness to experience (Selfhout et al., 2010).

We also observed self-monitoring, which represents the tendency of individuals to regulate their own behavior to create a desired impression and to control their self-presentation in social situations (Snyder, 1974). People with high self-monitoring rate tend to emphasize more their public appearance, while low self-monitors are more interested in their private realities (Snyder, 1987). Mehra, Kilduff, and Brass (2001) found that personality traits affect individuals' position in the social structure, and the longer people with high self-monitor scores work in an organization, the more they are likely to occupy positions of high betweenness centrality. Moreover, they show how personality, together with centrality, is predictive of the workplace performance of individuals. Dolgova (2013) recently found that high self-monitors are skillful in convincing teammates of their competence, which leads to establishing new friendship ties within the team.

### 2.3. The Creativity Factor

In this paper, we are interested in exploring how creativity influences the position of individuals in their friendship and advice networks. Creative ideas sow the seeds of successful innovation, therefore it is key to analyze the characteristics of the organizational context that can impede or support the generation of those ideas.

Creative individuals can generate new ways to perform their task, by coming up with novel and appropriate ideas (Perry-Smith and Shalley, 2003). Traits such as openness to experience, the degree of knowledge and individual intelligence are often cited as predictors of creativity (Jauk et al., 2014; Batey et al., 2010). For example, Wolfradt and Pretz (2001) found a positive relationship between openness to experience and several creativity measures, proposing that high scores in intuition and extraversion were the best predictors for creativity. Research in social psychology and leadership suggests that creative individuals differ in the way they approach work or engage with other people, favoring analytical rather than evaluative approaches and acting as gatekeepers (Mumford et al., 2002; Fiest and Gorman, 1998; Csikszentmihalyi, 1997).

While there is an extensive literature on how to recognize and nurture creative behaviors especially in organizations (Perry-Smith, 2006; Cattani and Ferriani, 2008; Fleming and Marx, 2006; Zhang et al., 2013), there is still poor empirical evidence regarding the correlation between creativity and ability to develop and maintain friendship and advice-seeking relationships with others. Combining many areas of specialized knowledge together requires the ability to connect several sources of knowledge, bridging interdisciplinary boundaries and leveraging weak ties. We would expect that a more central position of individuals in the advice network offers access to multiple perspectives that can stimulate brainstorming and the identification of innovative solutions.

### 2.4. Intelligence

To investigate the impact of intelligence on social network position, we adopt Cattell's concept of fluid intelligence, defined as the general ability to think abstractly, identify patterns, solve problems, and discern relationships (Cattell, 1943, 1963). Fluid intelligence is different from crystallized intelligence, which represents the individual acquired knowledge over time relating to specific information. Since we collected data on students' performance (final and average) at the end of the course and GPA, we considered performance as a partial proxy of crystallized intelligence. Previous studies have shown that the grade point average (GPA) could be used as an indication of students' intelligence and ability to perform well in a college setting (Hinds et al., 2000).

Intelligence has been often studied to predict academic performance (Spearman, 1904; Furnham and Zhang, 2005), improved outcomes in organizations (Schmidt and Hunter, 1998) and creativity (Guilford, 1942). Several studies suggest that intelligence is a strong predictor of job performance, particularly when task complexity increases (Judge et al., 2004; Gottfredson, 1997; O'Reilly and Chatman, 1994). Intelligence has been also suggested to be an important factor to determine leadership success (Stogdill, 1948; Mann, 1959). These studies seem to confirm that intelligence is an important antecedent of gaining social status and a central position in teams, organizations and informal networks. For example, homophily on traits like intelligence has been one of the first phenomena studied in the early social network literature (Almack, 1922). In this paper we go beyond the singular factor approach followed by many of these studies, since we explore how multiple factors impact the ability to create social networks. We also adopt a more dynamic and longitudinal perspective to the study of the predictors of centrality in social network.

### 2.5. Ability to Energize Others

Based on the research done by Casciaro and Lobo (2008) we define "energy" as the ability to inspire action and create a positive, pleasant drive in other people. Positive influence can encourage instrumental action, enabling access to information and resources relevant to the task, while negative affect can prevent important knowledge from being shared and used across the organization (Russell, 1980; Diener and Emmons, 1984). Affective traits such as the emotional contagion, or being inclined to catch and spread other people's emotions (Hatfield et al., 1994), can influence work behavior and impact organizational performance.

In their review of literature on affect in organizations, Barsade and Gibson (2007, p. 38) define "positive affectivity" as the tendency of individuals to be cheerful and energetic, and to experience positive moods, such as pleasure or well-being across a variety of situations, as compared to individuals who tend to be low energy and melancholy. In our study, we rely on this affect dimension – which is well-documented in the literature on intra-psychic mood - to investigate the social network position of people who have the ability to make others feel energized and up-beat during informal interactions.

It has been solidly demonstrated that affect is not only an intra-psychical trait, but has a strong social component able to influence the formation of dyadic and group interactions (Kelly and Barsade, 2001; Barsade and Gibson, 2007). Emotional contagion occurs when emotions get transferred from one individual to another, often without conscious knowledge of the individuals involved (Hatfield et al., 1994). Contagion has been demonstrated to be stronger in highly cohesive groups (Totterdell, et al., 1998), or for people with higher collectivistic tendency (Ilies, et al., 2007). Christakis and Fowler (2009) studied friendship, family, spousal, neighbour, and coworker relationships of 5124 individuals in the context of the Framingham Heart Study. They found that people who were surrounded by many happy people - and those who were also central in their social networks - were more likely to become happy in the future. Their study is particularly important as it shows how happiness spreads across a diverse array of social ties. Our study offers additional insights to understand how the ability to energize others fosters the development of social network in dyadic settings.

## 3. Research Model

Figure 1 presents the expected relations between popularity in both networks and independent variables. We used age and gender as control variables.

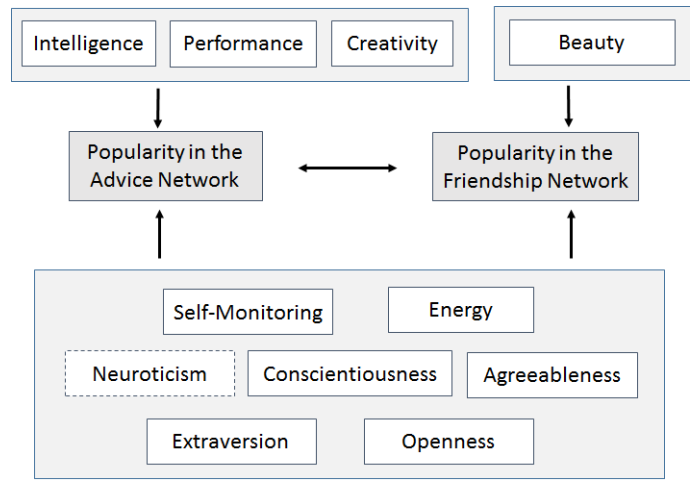

Figure 1. Expected relations between network popularity and independent variables (dotted lines for negative associations).

Based on the support from the literature described in the previous paragraphs, we propose the following hypotheses:

*H1. Personal attractiveness positively influences popularity in friendship networks.*

*H2. Extraversion, conscientiousness, agreeableness and openness to experience positively influence popularity in friendship and advice networks.*

*H3. Neuroticism negatively affects popularity in friendship and advice networks.*

*H4. Self-monitoring positively affects popularity in friendship and advice networks.*

*H5. The ability to energize others positively influences popularity in friendship and advice networks.*

*H6. Fluid intelligence, as well as perceived intelligence and good academic performance positively influence popularity in advice networks.*

*H7. Creativity and perceived creativity positively influence popularity in advice networks.*

Our hypotheses are meant to test the factors that affect peoples' popularity in the advice and friendship networks. Nonetheless, we include other effects in our models – described in details in the next paragraphs – to control for the other network dynamics and test the robustness of the popularity predictors. As previously discussed, we also control for any homophily effect that may arise due to individuals' similarities with respect to some of the above mentioned traits. At the same time, since we expect friendship ties to promote advice interactions and vice versa, we study the coevolution of the two networks to control for their reciprocal influence.

## 4. Experimental Setup

We collected data from October 2013 to January 2014, starting a month after the launch of a semester-long course on Production Plants held at a major university in Italy. The observation points are equally spaced between October, December 2013 and the end of the course in January 2014. In these three waves we mapped the advice and friendship networks of 197 BA students - 102 males and 95 females - who all responded to the network surveys. There were no students joining or leaving the class during the study. Students' performance was calculated based on the overall GPA prior to the course (initial cumulative performance) and on the final grade obtained at the end of the course (final performance).

In order to build the friendship network, we asked respondents to report the names of classmates with whom they felt they had developed meaningful ties outside the class and the context of the program. For advice-seeking ties we asked students to indicate the names of classmates they went to for advice on course-related topics or to solve a course-related problem following an approach similar to the one used by Snjders et al. (2013). Students were assured that their data would be kept confidential.

To collect data we used a name generator technique which has become the standard method in the study of personal or ego-networks to estimate social networks composition (Laumann, 1966). The limited size of the population (197 students) and the focus on the advice-seeking relations and meaningful ties among students should minimize issues of reliability and validity of this method. The name generator technique used in this study dealt with supportive exchanges that have specific criteria clearly defined in the survey which should make these generators less likely to be interpreted differentially across respondents (Marin and Hampton, 2007).

To measure the attractiveness/beauty factor, we asked students to express their level of agreement with the item "I think this person is attractive", per each other student they knew in the course (on a scale ranging from 1 to 5, where 1 is full disagreement and 5 is full agreement). Judgments were not anonymous, so we were able to build a dyadic variable represented via an adjacency matrix where cell $a_{ij}$ indicated the beauty score expressed by the student i with regard to the student j. We measured the variable "beauty" as a dyadic covariate. Accordingly, we included the main effect of the dyadic covariate to assess the influence of perceived beauty on advice and friendship interactions.

To assess personality profiles of students we referred to the five factor model of personality – widely used as general taxonomy (Judge et al., 2002) – and we administered a test based on the International Personality Item Pool (Goldberg et al., 2006). In particular, we used the 120-item version of the IPIP-NEO which covers the traditional five broad domains, namely neuroticism, extraversion, conscientiousness, agreeableness, and openness to experience. This test has demonstrated to be a reliable and valid measure of the five personality factors (Johnson, 2011, 2014). For each personality trait, the test provides both a numerical score – as percentile estimate – and an ordinal classification of the score (low/average/high), according to whether respondents of the same sex and age class scored approximately in the lowest 30%, middle 40%, or highest 30%.

To calculate self-monitoring, we used a version of the Self-Monitoring Scale measured on 18-true-false items (Snyder and Gangestad, 1986; Sullivan and Harnish, 1990). In our study the scale has been demonstrated to be reliable with a Cronbach's alpha of 0.74.

To measure creativity, we asked each student to find as many solutions as possible to a general shared problem in 15 minutes. Solutions were submitted in the form of mind maps representing diagrams to visually organize information: each node is a possible solution connected to the main problem raised by the instructors. Mind maps have been confirmed as a method to improve critical thinking, creativity and memory (Farrand et al., 2002). A creativity score was assigned to each student by two raters who evaluated the mind maps following a procedure which is similar to the one proposed by Torrance (1981). Similarly to the Torrance Test of Creative Thinking, they considered the number of responses, their originality and their level of detail (expressing a judgment on a scale which ranged from 1 to 5). Inter-rater agreement was fairly high (0.76) and was calculated by means of weighted Cohen's Kappa (Cohen, 1968).

To measure perceived creativity we asked students to anonymously express their level of agreement with the item "This person is creative and able to think outside of the box" rating the colleagues they knew, not only their friends (Casciaro and Lobo, 2008). Perceived Creativity is expressed on a 1 to 5 scale, where 1 indicates full disagreement and 5 full agreement.

To measure perceived intelligence we asked students to anonymously express their level of agreement with the item "I think this person is intelligent and competent". They were asked to rate the classmates they knew, including their close friends and acquaintances. The variable perceived intelligence was expressed on a 1 to 5 scale, where 1 is full disagreement and 5 is full agreement.

To test students' fluid intelligence we used the Raven Progressive Matrices Test (John and Raven, 2003). In the present research, Cronbach's alpha is 0.93. Fluid intelligence represents the ability to think abstractly and solve problems, independent of previous instruction concerning those relationships (Horn and Cattell, 1967; Cattell, 1963).

To monitor students' perceived energy we asked students to anonymously express their level of agreement with the item "When I interact with this person, I feel energized", rating the perceived level of energy of the colleagues they had an interaction before (using a 1 to 5 scale, where 1 is full disagreement and 5 is full agreement). This item aims to capture the activation dimension, using an approach that is consistent with Casciaro and Lobo (2008).

### 4.1. Models

In order to test the tendency of actors to send and receive ties and to explain their popularity, we built stochastic actor-oriented models for the dynamics and the co-evolution of one-mode networks; we use procedures and algorithms described by Snijders and colleagues (Ripley et al., 2014; Snijders et al., 2010). We employ the RSiena package of the statistical system R. In designing our model, we distinguish between endogenous structural effects (depending on the network structure itself), exogenous covariate-related effects (i.e. external variables with values that are not modeled, but are used to explain network change) and cross-network related effects, as described in the RSiena Manual. Covariates are exogenous variables used to explain network or behavior changes, whose values are not modeled (Ripley et al., 2014). The specified effects are representative of the way initial network structure and actors' covariates may affect network change. In order to control for network endogenous processes, we included in the models the structural effects described in Table 1.

Table 1. Description of structural effects (Ripley et al., 2014).

| Structural Effects | Definition |
| --- | --- |
| Out-degree | Expresses the balance between creating and deleting ties. |
| Reciprocity | The extent to which a tie from actor A to actor B leads to a tie from actor B to actor A. |
| Transitive triplets | The tendency for friends of friends to remain or become friends and similarly for advice relationships. For this effect the contribution of each tie is proportional to the total number of triplets it forms. |
| Three-cycles | The tendency for cyclical triadic closures in the network, i.e. if actor A is tied to B and B to C this leads to a tie from C to A. |
| Transitive ties | A tendency similar to the transitive triplets though it is not proportional to the total number of transitive triplets formed by a specific tie. |

| In-degree popularity | The tendency for actors with high values of in-degree to attract extra incoming ties because of their high in-degrees. This can be interpreted as the "rich get richer" effect (Price, 1976). |
|---|---|
| Out-degree popularity | The tendency for actors with high values of out-degree to attract extra incoming ties because of their high out-degree. |
| In-degree activity | The tendency for actors with high values of in-degree to send additional outgoing ties because of their high in-degrees. |
| Out-degree truncated | The tendency not to be an isolate node with respect to outgoing ties. This reflects the behavior of many actors who wish to nominate at least one of their colleagues as a friend or as somebody they contact for advices on the course topics. |
| Anti isolates | The tendency to connect to other social actors who otherwise would be isolate, disconnected, with zero incoming or outgoing ties. |

As a second step, we tested three main effects for all the monadic covariates, both in the advice and friendship network (Ripley et al., 2014). The first effect is called "*activity*", defined as the tendency of actors with a higher level of the covariate to increase more quickly the number of their outgoing ties (out-degree). The second effect is called "*popularity*", which is the tendency of actors with a higher level of the covariate to increase more quickly the number of their incoming ties (in-degree). The third effect is called "*similarity*", which can be regarded as homophily with respect to the covariate attribute. Similarity is defined as the tendency of actors to connect to people who have the same (or similar) values in the associated covariate. Examples of similarity include connecting to people of the same gender or with a similar score of creativity. Homophily suggests that "*distance in terms of social characteristics translates into network distance, the number of relationships through which a piece of information must travel to connect two individuals*" (McPherson et al., 2001, p.416).

Lastly, we analyzed the co-evolution of the advice and friendship networks, including the entrainment effect which assesses the influence of one network on the other (Ripley et al., 2014). Entrainment represents the extent to which the presence of a tie in one network promotes the creation of maintenance of a tie in the other network. This is in accordance with our hypothesis that being friends promotes advice interactions and vice versa.

Additional more complex effects from all the above mentioned categories were also tested without getting significant results. We followed an iterative process where almost all of non-significant effects were dropped from the models; only a few non-significant effects were retained because of their contribution to theory.

## 5. Results

The correlations results presented in Table 2 show that attractiveness is positively correlated with perceived intelligence, perceived creativity and extraversion, which reinforces the idea that sociable people are considered attractive people (Meier et al., 2010; Eagly et al., 1991), as well as competent and inventive. Only one of the personality traits, extraversion, is positively correlated with beauty. Extraverts usually enjoy social interactions and tend to be enthusiastic and gregarious, which could lead to transmit a positive image of themselves (Goldberg et al., 2006). This could also demonstrate that the core of the physical attractiveness stereotype is represented by factors such as popularity and sociability (Eagly et al, 1991).

Table 2. Pearson Correlations for Individual Attributes.

| | | Descriptive Statistics | | Correlation Coefficients | | | | | | | | | | | | | | |
|---|---|---|---|---|---|---|---|---|---|---|---|---|---|---|---|---|---|---|
| | | M | SD | 1 | 2 | 3 | 4 | 5 | 6 | 7 | 8 | 9 | 10 | 11 | 12 | 13 | 14 | 15 |
| 1 | Gender | 52% Males | | 1 | | | | | | | | | | | | | | |
| 2 | Age | 22.57 | 1.75 | .063 | 1 | | | | | | | | | | | | | |
| 3 | Fluid Intelligence | 52.91 | 6.61 | .021 | -.116 | 1 | | | | | | | | | | | | |
| 4 | Perceived Intelligence | 3.45 | 0.69 | .049 | -.119 | .189 | 1 | | | | | | | | | | | |
| 5 | Creativity | 2.91 | 1.34 | -.008 | -.081 | -.175 | -.037 | 1 | | | | | | | | | | |
| 6 | Perceived Creativity | 2.46 | 0.69 | .055 | .020 | .080 | .635** | -.107 | 1 | | | | | | | | | |
| 7 | Beauty | 2.88 | 0.77 | -.281** | .081 | -.042 | .450** | -.075 | .392** | 1 | | | | | | | | |
| 8 | Self-Monitoring | 7.56 | 3.51 | .262** | .080 | .047 | -.132 | .025 | -.095 | -.160 | 1 | | | | | | | |
| 9 | Extraversion | 50.85 | 23.72 | -.176 | -.066 | -.105 | -.017 | -.114 | .170 | .263* | .332** | 1 | | | | | | |
| 10 | Agreeableness | 33.88 | 21.06 | .027 | .184 | -.045 | -.057 | -.075 | .183 | .105 | -.265* | .037 | 1 | | | | | |
| 11 | Conscientiousness | 49.51 | 24.31 | .115 | -.215 | -.071 | .159 | -.193 | -.043 | .018 | -.031 | -.155 | .043 | 1 | | | | |
| 12 | Neuroticism | 42.21 | 22.15 | .024 | .031 | .131 | -.265* | .237 | -.102 | -.211 | .036 | -.235* | .031 | -.394** | 1 | | | |
| 13 | Openness to Experience | 39.40 | 25.38 | -.120 | .132 | -.102 | -.347** | .261 | -.215 | -.008 | .110 | .206 | .169 | -.111 | .083 | 1 | | |
| 14 | Average Performance | 23.89 | 1.52 | -.151* | -.341** | .276* | .362** | .055 | .100 | .002 | -.059 | -.041 | -.189 | .244* | -.030 | -.339** | 1 | |
| 15 | Final Performance | 24.39 | 3.76 | .104 | -.117 | .188 | .256* | -.041 | .142 | .098 | .157 | .132 | -.229 | -.042 | -.094 | -.291* | .295** | 1 |

\* p < 0.05; ** p < 0.01.

Both final performance and average performance (GPA) were positively correlated with perceived intelligence. Performance was negatively correlated with openness to experience which seems to contrast previous studies suggesting that openness is positively correlated with learning motivation, approach to learning and critical thinking (Vermetten et al., 2001; Tempelaar et al., 2007; Bidjerano and Dai, 2007). A possible reason for this negative correlation is the nature of the work that students had to perform in this class, or the expectations set by instructors to perform well, both in the course and in other courses within the Engineering program. Most engineering teachers tend to teach deductively, based on lecture model. This could lead students - who are less inclined to think outside the box – to get a better grade by carefully following the assignments instructions and by not deviating from the suggested path.

Neuroticism was negatively correlated to conscientiousness, perceived intelligence and extraversion, which is aligned with previous studies (Costa and McCrae, 1992; Larsen and Ketelaar, 1989). As expected, we see that attractiveness (beauty) was positively correlated with perceived intelligence and creativity: probably as a consequence of the "halo effect" (Verhulst et al., 2010; Thorndike, 1920).

Stability between consecutive observations of networks is required for a meaningful application of the SIENA method. Stability can be measured by the Jaccard index as suggested by Snijders et al. (2010); coefficients are fairly good for our networks (values above .6 are preferable and between .3 and .6 acceptable) – as shown in Table 3.

Table 3. Networks descriptive statistics and Jaccard Coefficients.

|  | Fr. T1 | Fr. T2 | Fr. T3 | Adv. T1 | Adv. T2 | Adv. T3 |
|---|---|---|---|---|---|---|
| Av. Degree | 5.482 | 5.629 | 5.208 | 0.904 | 1.269 | 1.147 |
| Reciprocity | 0.711 | 0.711 | 0.673 | 0.360 | 0.368 | 0.338 |
| Clustering | 0.454 | 0.483 | 0.462 | 0.315 | 0.324 | 0.312 |
|  |  |  |  |  |  |  |
|  |  | Fr. T1 to T2 | Fr. T2 to T3 |  | Adv. T1 to T2 | Adv. T2 to T3 |
| Jaccard Index |  | 0.802 | 0.790 |  | 0.523 | 0.649 |

Average degree is much higher for the friendship network and it increases after the first wave in the advice network. A possible reason could be that first data collection was closer to the beginning of the course and students did not have enough time to build more stable relationships as they did for friendship (a subset of students knew each other from previous classes). This is also reflected in the Jaccard index which shows a higher stability for the friendship network. Furthermore, course topics tend to be easier in the first few weeks, which could lead to less need for peer support.

As expected, the friendship network shows a stronger tendency toward transitivity. Both networks show a tendency toward reciprocity, again stronger for the friendship network. We used the quadratic assignment procedure (QAP) to calculate network correlations and we tested their significance considering distributions generated from 50,000 random permutations (Krackhardt, 1988; Noreen, 1989). Correlations are all significant and fairly high (see Table 4); they increase from first to second wave of observation, indicating that as students get more time to socialize, they strengthen their ties and are more likely to extend their advice network based on friendship, familiarity and trust.

Table 4. QAP correlations of friendship and advice networks.

|  | T1 | T2 | T3 |
|---|---|---|---|
|  | Friendship | Friendship | Friendship |
| Advice | 0.383* | 0.459* | 0.456* |
| * $p < 0.001$ | | | |

Both friendship and advice networks exert a strong influence on one another, with friendship being more influential on the advice network than the other way around. The models show a good convergence (having for all the parameters a t-ratio smaller than 0.1). Table 5 summarizes the final model results. Fit statistics for final models are good with respect to in degrees, out degrees, triad census and geodesics distributions, both for friendship and advice networks.

With regard to the structural effects, we notice a tendency in both networks to reciprocate the social ties and a tendency toward closure in a more hierarchical way. These results are in line with previous research (Snijders et al., 2013). The "rich get richer" effect - represented by the in-degree popularity - is only significant in the friendship network and is negative, which indicates that students prefer to connect to less popular peers. This seems to be aligned to Gould's theory (2002) that explains why winner-take-all hierarchies are discouraged. As Gould explains, "*the displeasure of offering unreciprocated gestures of approval keeps such gestures within limits*". (Gould, 2002, p.1149). This result is also reinforced by the positive "anti isolates" effect. In the friendship network people who are nominated by many colleagues as friends seem less willing to send out new ties; in general we notice that many actors tend to nominate at least one friend in the course. Lastly, in the advice network we observe a negative out-degree popularity effect. This is probably due to the fact that students who ask for advice from many others are considered less reliable sources of information. A tendency towards homophily is significant in both networks only with respect to students age and perceived energy, not with respect to all the other covariates. An important difference between the friendship and the advice networks is that friendship is more strongly transitive and has a significant anti-isolate effect, which indicates the tendency to connect to other students who otherwise would be disconnected.

Table 5. Models results

|  | Friendship | | Advice | |
|---|---|---|---|---|
| Effect | Par. | (s.e.) | Par. | (s.e.) |
| Out-degree | -2.157*** | (0.381) | -6.916*** | (0.376) |
| Reciprocity | 3.005*** | (0.355) | 1.975*** | (0.374) |
| Transitive triplets | 1.720*** | (0.292) | 0.691*** | (0.167) |
| Three-cycles | -1.090*** | (0.212) | -0.365 | (0.306) |
| Transitive Ties | 1.553*** | (0.221) | -- | -- |
| In-degree popularity | -0.075** | (0.023) | -0.013 | (0.069) |
| Out-degree popularity | -- | -- | -0.683** | (0.212) |
| In-degree activity | -0.428*** | (0.097) | -- | -- |
| Out-degree truncated | -3.870*** | (0.543) | -- | -- |
| Anti isolates | 1.669*** | (0.454) | -- | -- |
| Beauty | 0.451** | (0.172) | 0.432*** | (0.126) |
| Same Gender | 0.015 | (0.143) | -- | -- |
| Age similarity | 2.035** | (0.785) | 4.538** | (1.586) |
| Creativity popularity | -- | -- | 0.206* | (0.100) |
| Creativity activity | -- | -- | 0.287* | (0.126) |
| Perceived intelligence popularity | -- | -- | 0.739*** | (0.209) |
| Perceived creativity popularity | -- | -- | 0.708** | (0.245) |
| Perceived energy popularity | -- | -- | -0.657** | (0.225) |
| Perceived energy activity | -- | -- | 1.116*** | (0.223) |
| Perceived energy similarity | 2.479*** | (0.718) | 1.933* | (0.925) |
| Average Performance popularity | -- | -- | 0.259** | (0.083) |
| Self-monitoring activity | 0.089*** | (0.024) | 0.093** | (0.035) |
| Neuroticism activity | -- | -- | 0.020** | (0.007) |
| Openness to Experience popularity | -- | -- | 0.013* | (0.005) |
| Openness to Experience activity | -- | -- | 0.017* | (0.007) |
| Advice | 2.998*** | (0.849) | -- | -- |
| Friendship | -- | -- | 5.101*** | (0.371) |
| * $p < 0.05$; ** $p < 0.01$; *** $p < 0.001$. | | | | |

In support for H1, we found that beauty plays an important role in increasing the interactions in both networks, thus including the advice network. These results seem to indicate that beauty is one of the first determinants that will make someone popular in their circles, leading to more friends and more advice interactions. These results are aligned to Webster and Driskell's work (1983) who found that beauty produces ideas of task competencies, which could explain the positive association between physical attractiveness and advice network centrality. In the friendship network we do not find other covariates that significantly influence popularity neither in personality traits, nor in other traits. We only find that actors with higher scores of self-monitoring are more active in sending out ties. Consequently, H2, H3 and H4 are not supported for the friendship network. People with high self-monitoring score may think more strategically about their advice ties than they do about their friendship ties, which might help explain why some associations, such as openness to experience, only hold for instrumental networks. Surprisingly, we did not find support for H3 also in the advice network. Though neuroticism has been empirically related to loneliness, to anxiety, jealousy and a sense of vulnerability (Costa and McCrae, 1992; Russell et al., 1997), our results do not indicate that it also negatively affects the ability to make friends or to be sought out for course-related questions. On the other hand, being popular in the advice network is not only positively influenced by beauty, but also by measured creativity, perceived creativity, perceived intelligence, average past performance and by being more open to experience which is a trait often associated with creativity (Kaufman et al., 2015; Jauk et al., 2014). Measured fluid intelligence does not play a role in both networks. These findings fully confirm H7 and partially support H6 (only fluid intelligence is not significant) and H2 (only with regard to openness to experience).

Moreover, the results show that students who share a similar level of energizing skills connect more with each other. Higher scores of perceived energy seem to hinder popularity in the advice network – H5 is not supported both for friendship and advice networks. In addition, the more a student is perceived as someone who transfers high energy, the less likely he/she is to be sought out for a task interaction, yet he/she is more likely to be active establishing new ties (higher out-degree). Lastly, more active actors on the advice network are those with higher levels of openness to experience and measured creativity, probably because they look for more diverse information sources to integrate; neuroticism, since more anxious students might want to ask for more advices; perceived energy; self-monitoring, as it happens for the friendship network.

To sum up, we find full support for H1 and H7; we find support for H2 only with regard to openness to experience in the advice network; we find no support for H3; we find partial support for H6, only in the advice networks excluding the non-significant effect of fluid intelligence; we find no support for H4 since self-monitoring is not associated with a significant popularity effect in any of the networks. As regards H5, we find no evidence in the friendship network, whereas we notice that, in the advice network, being perceived as more energizing might hinder (instead of foster) the request for advices.

6. **Discussion**

In this study we were interested in finding the main determinants of popularity by exploring the impact of personal attractiveness (H1), personality traits (H2, H3 and H4), ability to energize others (H5), intelligence (H6) and creativity (H7). Figure 2 illustrates the actual relations between variables. Along with the surprising connection between beauty and advice network, we found no significant relationship between popularity in the advice network and extraversion, neuroticism, self-monitoring, agreeableness or conscientiousness. This seems to be aligned with other studies on personality traits and social network structure suggesting that the relationship between personality and network characteristics is more complex than previous studies have indicated (Pollet et al., 2011; Mehra et al., 2001).

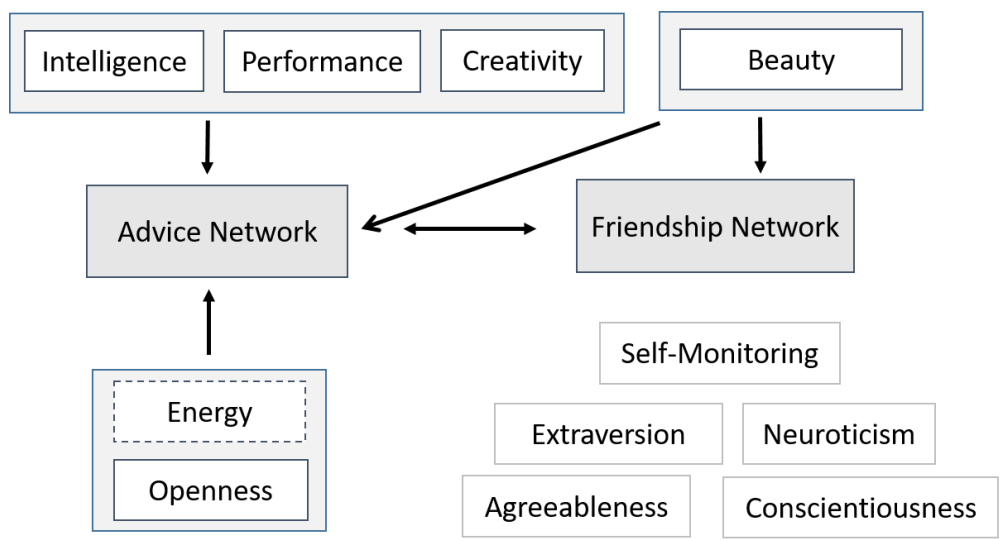

Figure 2. Determinants of popularity (dotted lines for negative associations).

Our findings support the idea that physical attractiveness and a shared ability to energize others are two complementary factors that influence popularity in advice networks. Similarly to other studies on emotional contagion, we found that individuals will be attracted by others whom they consider good-looking and with whom they share a similar level of energy (Barsade and Gibson, 2007; Totterdell et al., 1998; Snijders et al., 2013).

Our study proves that physical attractiveness counts not only towards the ability to attract more friends, but it also influences the individual's chance to be considered for task related information. It appears that people prefer to seek advice from individuals they like who are pleasant, more sociable, and not necessarily more competent (Meier et al., 2010; Eagly et al., 1991). This result is aligned to and reinforces the conclusions presented by Casciaro and Lobo (2008) who suggested the possibility that "*the way people form work relationships may follow laws of social interaction that are not highly susceptible to contingent characteristics of the task*" (Casciaro and Lobo, 2008, p.678).

On average, being liked seems to be as important as being considered intelligent or highly performance when it comes to be sought out for a task interaction. While Casciaro and Lobo (2008) do not explicitly investigate the impact of physical attractiveness and focus on "pleasantness" and "liking", we found that the way people look plays a key role in determining the attribution of competence. Our findings on the role of physical attractiveness and ability to energize others contribute to the psychological literature on emotions in organizations (Barsade and Gibson, 2007; Brief and Weiss, 2002). Besides corroborating this halo effect linking interpersonal traits and beauty, our study suggests that beautiful people are asked more often for help, probably because "*their perceivers may infer on the basis of attractive people's presumed social competence that they have other favorable traits*" (Eagly et al, 1991, p. 112). The mental association between beauty and nonphysical traits could be based on a direct observations of attractive and less attractive people in their social environment. Furthermore, literature on the role of affect in organizations suggests that physical attractiveness is aesthetically pleasing and therefore may stimulate positive affect that leads others to ascribe favorable characteristics to attractive people (Barsade and Gibson, 2007; Brief and Weiss, 2002).

In addition to stress the role of beauty, our findings highlight that perceived intelligence, creativity (both measured and perceived) and openness to experience are other antecedents of popularity in the advice network. This result is aligned with other studies showing how highly creative scientists are engaged in extensive communication with peers and act more often as gatekeepers (Mumford et al., 2002).

Students with a higher GPA are not surprisingly more popular, as they are asked for advice more frequently based on an attribution of expertise and competence. Creative students, as well as the ones with high openness to experience, seem to be active and popular in the advice network, but not in the friendship network. Students who are perceived as more energizing do not seem to attract peers when it comes to asking for academic advice, though they are active trying to get others to help with course-related problems. Homophily, that is the tendency to connect to people who have similar values, seems to be a very important factor in determining friendship and advice-seeking ties for people with the ability to energize others.

With this research we try to overcome the main limitation of previous studies, namely their singular and mono-dimensional approach. Previous research has focused on the impact of personality traits and self-monitoring attitude on centrality (Fang et al., 2015; Klein et al., 2004; Mehra et al., 2001; Pollett et al., 2011); others emphasized the role of creativity (Cattani and Ferriani, 2008; Fleming and Marx, 2006) or intelligence (Jeung, 2013). Our findings extend existing research on determinants of social network position and popularity by focusing on the role of multiple factors including physical attractiveness, personality traits, creativity, intelligence, performance, ability to energize others and self-monitoring.

Even though the setting of this study was represented by an academic environment, we believe our findings might be extended to other settings where individuals work both independently and within project teams. For example, studying the determinants of popularity has important managerial implications for entrepreneurs whose success depend on the ability to create and nurture new business contacts.

## 7. Study Limitations

In this study we have analyzed the effects of personality values, intelligence, energy level, creativity and beauty on individual popularity. We want to recognize that other predictors may also play an important role in the advice and friendship network, which requires further investigation. These factors include connections outside of the classroom environment, such as membership to student associations or romantic relationships. The sample of 197 students included people who had attended previous courses together, so it would be key to differentiate between newly established and old friendship ties (a distinction we could not make in our sample).

Another limitation, as well as an opportunity to further this research, is the lack of data on the students' socio-economic status due to institutional constraints. Because of the powerful role that an individual's or family's economic and social status play in determining social network position (McPherson et al., 2001; Lin, 1999), we suggest that any replication of this study includes collecting socio-demographic information from participants.

Lastly, different courses – focusing on different topics and thus attracting students other than future engineers – could be analyzed, together with a longer research timeframe, to see if our findings are replicated.

Krantz, M. (1987) 'Physical Attractiveness and Popularity: A Predictive Study', *Psychological Reports*, 60(3), pp. 723–726. doi: 10.2466/pr0.1987.60.3.723.

Krebs, D. and Adinolfi, A.A. (1975) 'Physical attractiveness, social relations, and personality style', *Journal of Personality and Social Psychology*, 31(2), pp. 245–253. doi: 10.1037/h0076286.

Kubitschek, W.N. and Hallinan, M.T. (1998) 'Tracking and students' friendships', *Social Psychology Quarterly*, 61(1), p. 1. doi: 10.2307/2787054.

Langlois, J.H., Kalakanis, L., Rubenstein, A.J., Larson, A., Hallam, M. and Smoot, M. (2000) 'Maxims or myths of beauty? A meta-analytic and theoretical review', *Psychological Bulletin*, 126(3), pp. 390–423. doi: 10.1037/0033-2909.126.3.390.

Larsen, R.J. and Ketelaar, T. (1989) 'Extraversion, neuroticism and susceptibility to positive and negative mood induction procedures', *Personality and Individual Differences*, 10(12), pp. 1221–1228. doi: 10.1016/0191-8869(89)90233-x.

Laumann, E. (1966). *Prestige and Association in an Urban Community*. Bobbs-Merrill, New York.

Lin, N. (1999) 'Social Networks and Status Attainment', *Annual Review of Sociology*, 25(1), pp. 467–487. doi: 10.1146/annurev.soc.25.1.467.

Mann, R.D. (1959) 'A review of the relationships between personality and performance in small groups', *Psychological Bulletin*, 56(4), pp. 241–270. doi: 10.1037/h0044587.
43